\documentclass[aps,prl,twocolumn,showpacs,superscriptaddress,preprintnumbers,floatfix,amsmath,amssymb,eufrak,table]{revtex4-1}
\pdfoutput=1
\usepackage{graphicx}
\usepackage{dcolumn}
\usepackage{bm}
\usepackage{slashed}
\usepackage{amsmath,graphicx}
\usepackage[colorlinks=true,linktocpage=true,linkcolor=blue,citecolor=blue]{hyperref}
\usepackage{float}
\usepackage{nicefrac}
\usepackage[normalem]{ulem}
\usepackage{amsmath}
\usepackage{subfigure} 
\usepackage{bbold} 
\usepackage[makeroom]{cancel}
\usepackage{stmaryrd} % for weird double brackets
\begin{document}
\title{Open charm phenomenology with a multi-stage approach to relativistic heavy-ion collisions}

\author{Mayank Singh}
\email{singh547@umn.edu}
\affiliation{School of Physics \& Astronomy, University of Minnesota, Minneapolis, MN 55455, USA}

\author{Manu Kurian}
\email{mkurian@bnl.gov}
\affiliation{Department of Physics, McGill University, 3600 University Street, Montreal, QC, H3A 2T8, Canada}
\affiliation{RIKEN BNL Research Center, Brookhaven National Laboratory, Upton, New York 11973, USA}

\author{Sangyong Jeon}
\email{jeon@physics.mcgill.ca }
\affiliation{Department of Physics, McGill University, 3600 University Street, Montreal, QC, H3A 2T8, Canada}

\author{Charles Gale}
\email{gale@physics.mcgill.ca }
\affiliation{Department of Physics, McGill University, 3600 University Street, Montreal, QC, H3A 2T8, Canada}

%%%%%%%%%%%%%%%%%%%%%%%%%%%%%%%%%%%%%%%%%%%%%%%%%%%%%%%%%%%%%%%%%%%

\begin{abstract}

 We study open charm flavor observables in Pb+Pb collision at $\sqrt{s_{NN}}= 2.76$ TeV within the MARTINI framework. The space-time expansion of the quark-gluon plasma is described using the hydrodynamical approach-MUSIC with IP-Glasma initial conditions. The model parameters, including the viscous coefficients, were obtained from a recent Bayesian model-to-data comparison. We evolve heavy quarks in this background using Langevin dynamics while incorporating their collisional and radiative processes in the medium. The sensitivity of charm observables to the IP-Glasma initial state,  bulk evolution, and centrality of the collision is studied. We find that the elliptic flow of open charm flavor has a strong dependence on the fluctuating initial conditions in addition to the strength of the interaction of heavy quarks with the medium constituents. Within this framework, the nuclear suppression factor and elliptic flow of D-mesons act as efficient probes to study the initial stages of heavy-ion collisions, transport coefficients associated with QGP medium as well as heavy quark interactions. 

\end{abstract}

%%%%%%%%%%%%%%%%%%%%%%%%%%%%%%%%%%%%%%%%%%%%%%%%%%%%%%%%%%%%%%%%%%%%

\maketitle
%%%%%%%%%%%%%%%%%%%%%%%%%%%%%%%%%%%%%%%%%%%%%%%%%%%%%%%%%%%%%%%%%%%%
%%%%%%%%%%%%%%%%%%%%%%%%%%%%%%%%%%%%%%%%%%%%%%%%%%%%%%%%%%%%%%%%%%%%
{\bf \emph{Introduction}}- Heavy-ion collision experiments at the Relativistic Heavy Ion Collider (RHIC) and Large Hadron Collider (LHC) provide an access to strongly interacting matter: the Quark-Gluon Plasma (QGP). Relativistic viscous hydrodynamics serves as an important framework to study the dynamical evolution of the QGP\cite{Gale:2013da}.
The transport properties of QGP are extracted from model to light flavor hadron data comparisons, and significant effort has been devoted in this direction\cite{Romatschke:2007mq,Heinz:2013th, Ryu:2015vwa,Jaiswal:2016hex}. Recently, Bayesian analysis has been used to do a systematic extraction of shear and bulk viscosities of the medium and their uncertainty quantification ~\cite{Bernhard:2019bmu,JETSCAPE:2020shq,Nijs:2020roc,Heffernan:2023utr}.

Heavy flavor quarks provide additional probes to study the properties of QGP~\cite{vanHees:2005wb,Das:2009vy,He:2012df,Andronic:2015wma,Aarts:2016hap,Song:2019cqz,Mustafa:2004dr,Sun:2019fud}. They are largely created in the initial stages of the collision, and pass through the QGP while interacting with the light flavor quarks and gluons. Thermal production of heavy quarks in the QGP medium is expected to be negligible because of their larger mass compared to the temperature scale of the medium~\cite{Dong:2019unq}. Generally, heavy flavor particles are not treated as part of the medium as the thermalization time is longer than the QGP lifetime. Their dynamics can be described as Brownian motion in the medium and can be studied within the Langevin or the Fokker-Planck frameworks~\cite{GolamMustafa:1997id,Moore:2004tg,Uphoff:2011ad,Das:2013kea,Cao:2016gvr,Li:2019wri}. Nuclear suppression factor $R_{AA}$ and elliptic flow $v_2$  are the key observables associated with the heavy flavor particles at the RHIC and LHC energies. Several studies have been done to estimate $R_{AA}$ and $v_2$ by modeling the Brownian motion of the heavy quarks in the medium~\cite{Akamatsu:2008ge,He:2011qa,Cao:2013ita,Alberico:2013bza,Song:2015sfa,Li:2018izm,vanHees:2007me} and to extract the momentum and temperature behavior of heavy quark transport coefficients from these observables in heavy-ion collision experiments~\cite{Xu:2017obm,Scardina:2017ipo,Rapp:2018qla,Cao:2018ews}. Notably, the inclusion of inelastic interactions of heavy quarks in the medium on top of the elastic collisions reduces the gap between experimental and theoretical observations~\cite{Wicks:2005gt}.  

The evolution history of the collision event is essential to model the heavy quark dynamics in the expanding medium. Heavy quark production and transport have been widely studied in the static limit and in expanding fireball models. Some efforts have been made to study the heavy flavor dynamics in evolving medium within 1+1 Bjorken hydrodynamics as well as higher dimension hydrodynamics~\cite{Young:2011ug,He:2013zua,Sarkar:2018erq,Thakur:2020ifi,Prakash:2021lwt}. Most analyses utilized smooth initial conditions for the hydrodynamical evolution of the QGP medium through which heavy quarks traverse. Advances in hydrodynamical models to describe medium expansion seem to have a significant impact on the heavy quark observables~\cite{Gossiaux:2011ea,Song:2020tfm,JETSCAPE:2022hcb}. It is also known that the initial state fluctuation has a significant influence on the light flavor jet energy loss~\cite{Rodriguez:2010di,Zhang:2012ik}. 

In this work, we employ the state-of-the-art hydrodynamical model of QGP to study the charm observables. We consider the evolution history of a Pb+Pb collision event at 2.76 TeV energy with the IP-Glasma initial state~\cite{Schenke:2012wb,Schenke:2012hg,McDonald:2016vlt}. IP-Glasma is a very successful dynamical model which describes a variety of observables in heavy-ion collisions. The initial fluctuating color configuration in the heavy-ion nuclei that are approaching at high velocity can be determined within IP-Sat approach~\cite{Bartels:2002cj,Kowalski:2003hm} combined with the Yang-Mills equations~\cite{Krasnitz:1998ns}. Its evolution also follows from solving the Classical Yang-Mills equations. The Glasma distributions, which are obtained event-by-event, act as the input for the hydrodynamical evolution. We used the recently updated shear and bulk viscous coefficients obtained from the Bayesian model-to-data analysis~\cite{Heffernan:2023utr}.

The charm dynamics are encoded in the drag and diffusion coefficients in the Langevin equation. The Langevin equation is solved within MARTINI~\cite{Schenke:2009gb,Young:2011ug}. We studied open charm observables within three different setups of drag and diffusion coefficients. This analysis is an up-to-date study of the heavy flavor nuclear suppression factor and elliptic flow using the recent developments in the hydrodynamical description of QGP and drag and diffusion coefficients of the charm quarks.

%%%%%%%%%%%%%%%%%%%%%%%%%%%%%%%%%%%%%%%%%%%%%%
%%%%%%%%%%%%%%%%%%%%%%%%%%%%%%%%%%%%%%%%%%%%%%
{\bf \emph{A hybrid model for heavy quarks}}- Modelling of heavy flavor evolution in collisions can be divided into three distinct stages: heavy quark initial production, its evolution in the hydrodynamized expanding medium, and the hadronization process. In the current analysis, we employ PYTHIA8.2~\cite{Sjostrand:2014zea} for perturbative production of charm quarks by sampling 6-dimensional momentum distributions of $Q\bar{Q}$ systems. We allow the gluons to split into $c\bar{c}$ pairs, but the medium-induced modification of the gluon splitting rate is not accounted for. As the nucleons are bound in a heavy nucleus, the parton distribution functions (nPDFs) will be modified. To take into account the nuclear shadowing effect, EPS09 nuclear parton distribution functions are used~\cite{Eskola:2009uj}. Isospin effects are accounted for by sampling $p+p$, $p+n$ and $n+n$ collisions. A finite thermalization time $\tau_0$ in the heavy-ion collision is considered, and the evolution of charm quarks during this time is governed by the equation of motion with the zero-temperature Cornell potential~\cite{Young:2011ug}.

The QGP medium is initialized using the IP-Glasma model~\cite{Schenke:2012wb,Schenke:2012hg,McDonald:2016vlt} and evolved using the viscous hydrodynamical approach MUSIC~\cite{Schenke:2010nt,Schenke:2010rr}. The lattice QCD equation of state (EoS) from the hotQCD collaboration~\cite{HotQCD:2014kol} at high temperature is smoothly matched to the hadron resonance gas EoS at low temperatures and is incorporated in the framework. The first thorough Bayesian model-to-data comparison of relativistic heavy-ion collision measurements using a hybrid model combining viscous expansion (MUSIC), evolution to particlization (iS3D)~\cite{McNelis:2019auj}, and particle transport (SMASH)~\cite{Weil:2016zrk} with IP-Glasma initial conditions has been recently presented for four different model choices in~\cite{Heffernan:2023utr}. We chose the model with Grad's 14-moment viscous correction with constant shear viscosity to entropy density ratio and a temperature-dependent bulk viscosity profile. In the present analysis, we have used the maximum a posteriori estimates of all model parameters, including the viscous coefficients, from this study.

The dynamics of heavy quarks in the QGP medium depend upon its radiative and elastic interactions with the constituents in the medium. The strength of interactions of heavy quarks in the medium can be quantified in terms of drag and diffusion coefficients. In the local rest frame of the medium, the heavy quark motion can be studied numerically using the discrete version of the Langevin equations~\cite{Moore:2004tg,Das:2013kea},
\begin{align}
   dp_i=-A_i\, dt+C_{ij}\rho_j\sqrt{dt},
\end{align}
where $dp_i$ denotes the change in momentum in time interval $dt$. The drag force $A_i$ and covariance matrix $C_{ij}$ are defined as follows,
\begin{align}\label{eq:A}
    &A_i=p_iA(|{\bf p}|^2, T),\\\label{eq:B}
    &C_{ij} = \sqrt{2B_0}\left(\delta_{ij}-\frac{p_ip_j}{|{\bf p}|^2}\right)+\sqrt{2B_1}\frac{p_ip_j}{|{\bf p}|^2},
\end{align}
where $A$ denotes the drag coefficient and the quantities $B_0$ and $B_1$ represent the transverse and longitudinal momentum diffusion coefficients, respectively. The drag force describes the heavy quark average momentum transfer due to the interactions, whereas the matrix $C_{ij}$ quantifies the stochastic force by using Gaussian-normal distributed random variable $\rho_j$\cite{Das:2013kea}. The dependence of drag and diffusion on heavy quark momentum and temperature of the medium can be studied within relativistic transport theory. The Langevin dynamics of the heavy quarks is coupled with the expanding QGP medium and solved within MARTINI as follows:
\begin{itemize}
    \item[$-$] Find the fluid four-velocity and temperature at the space-time location of the charm quark from MUSIC
    \item[$-$] Boost the charm momentum to the fluid local rest frame and evolve the charm three momenta to the next time step using Langevin equations 
    \item[$-$] Boost the charm momentum back to the lab frame and update the charm position after the time step 
\end{itemize}
The heavy quark transport coefficients are the key input parameters that quantify the interaction strength of heavy quarks with the medium. The transport coefficients can be derived in perturbative QCD by including scattering and radiative processes~\cite{Svetitsky:1987gq,Mustafa:2004dr}. The specific interactions are encoded in the matrix elements. It has been seen that using Debye mass as infrared (IR) regulator ($\mu_{IR}$) in the gluon propagator for t-channel interaction and a fixed coupling constant, pQCD matrix elements are not able to describe the experimental data associated with the heavy flavor particles~\cite{Rapp:2009my,Rapp:2018qla}. As these parameters are the sources of uncertainty in the estimation of heavy quark transport coefficients, the IR regulator and coupling constant are determined in the analysis by physical arguments as described in~\cite{Gossiaux:2008jv}.  In the present analysis, the conventional choice of IR regulator, the Debye mass, is replaced with a realistic hard thermal loop (HTL) parameterization of the IR regulator. Further, an effective coupling constant ($\alpha_{\text{eff}}$) that embeds non-perturbative dynamics is employed in the study.  The behaviour of $\alpha_{\text{eff}}$ is obtained from the analysis of $e^+e^-$ annihilation~\cite{Mattingly:1993ej} and decay of $\tau$ leptons~\cite{Brodsky:2002nb}, and has the following form~\cite{Gossiaux:2008jv},
\begin{equation}
    \alpha_{\text{eff}}(R^2) = \dfrac{4\pi}{\beta_0}\Bigg\{
    \begin{array}{lr}
    L_-^{-1}, &  R^2<0\\
    \frac{1}{2}-\pi^{-1}\text{arctan} (L_+/\pi), & R^2>0
    \end{array}
\end{equation}
where $R$ is the relevant energy scale, $\beta_0=11-\frac{2}{3}N_f$ with $N_f$ as the number of flavors and $L_{\pm}=\ln{(\pm R^2/\Lambda^2)}$ with QCD parameter chosen as $\Lambda=0.2$ GeV. With these parameterizations, the propagator with bare coupling $\alpha_s$ is modified for $t-$channel gluon exchange processes (heavy quark-thermal medium interaction) as,
\begin{align}
    \frac{\alpha}{t}\rightarrow \frac{\alpha_{\text{eff}}\,(t)}{t-\mu^2_{IR} (t, T)},
\end{align}
where $t$ is the Mandelstam variable and $\mu^2_{IR}(t, T)=\kappa\, 4\pi \Big(1+\frac{N_f}{6}\Big)\alpha_{\text{eff}}\,(t) T^2$ with $\kappa=0.2$. IR regulator is not required for other channels and coupling constant is fixed as $\alpha\rightarrow \alpha_{\text{eff}}\,(s-m^2_{HQ}) $ and $\alpha\rightarrow \alpha_{\text{eff}}\,(u-m^2_{HQ}) $ for $s$ and $u-$ channels, respectively such that $s=m^2_{HQ}$ and $u=m^2_{HQ}$ denote the maximal softness in the channels~\cite{Gossiaux:2008jv}. The matrix elements $ {\mathcal{M}}_{2\rightarrow 2}$ and  $ {\mathcal{M}}_{2\rightarrow 3}$ (that appear on eqs.~\ref{Ap1c} and and~\ref{Ap1}) quantify the interaction strength of elastic and inelastic process processes of charm quark in the medium. The modified IR regulator (for $t-$channel processes) and effective coupling enter into the estimation of charm quark transport coefficients through these matrix elements. Here, $m_{HQ}$ is the mass of the heavy quark. For the charm quark, we took $m_{HQ} = 1.25$ GeV.
\begin{figure}
    \includegraphics[width=0.45\textwidth]{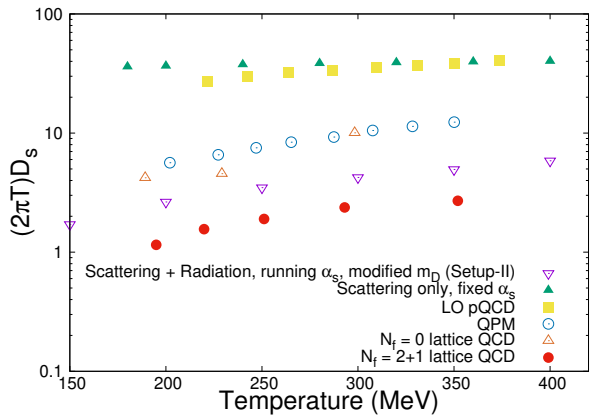}
    \caption{Temperature dependence of spatial diffusion coefficient of charm quark and comparison of the results with other approaches.}
    \label{fig1}
\end{figure}

In this study, we consider three different setups to characterize the coupling strength of heavy quarks in the QGP medium. {\it (i) Setup I-Constant value of $2\pi D_sT$:} In the static limit of heavy quarks, the coupling of heavy quarks in the QGP can be described with one physical parameter-the spatial diffusion coefficient $D_s$, which is often treated as a phenomenological parameter. For $\sqrt{s_{NN}} = 2.76$ TeV energy collisions, we take $2\pi D_sT=3.0$. {\it (ii) Setup II-Temperature dependent heavy quark transport coefficients:}  Employing the fluctuation-dissipation theorem, $D_s$ can be expressed in terms of a drag coefficient in the limit $p\rightarrow 0$ as, $D_s=\frac{T}{m_{HQ}A(p\rightarrow 0,T)}$. The momentum dependence of $D_s$ is neglected in this scenario. The temperature dependence of $D_s$ for the collisional and radiative processes with the effective coupling and modified IR regulator is depicted in fig.~\ref{fig1}. For the pQCD elastic collisional process, the value of $(2\pi T)D_s$ lies in the range of $30-40$~\cite{vanHees:2004gq,Kurian:2020orp}, which is an order of magnitude larger than that obtained from $N_f=0$ lattice result~\cite{Banerjee:2011ra}, $N_f=2+1$ lattice data~\cite{Altenkort:2023oms} and quasiparticle model (QPM)~\cite{Scardina:2017ipo} estimation. It is seen that the gluon radiation of heavy quarks in the medium suppresses the $D_s$. This can be understood from the fact that the heavy quark is experiencing more drag as they lose energy through collisional and radiative processes while traveling through the QGP medium. The effective coupling that incorporates the non-perturbative effects and the HTL IR regulator seems to have significant impacts on the temperature behavior of $D_s$. With the above choice of parameters and the inclusion of the radiative process in the analysis, it is observed that $(2\pi T)D_s\approx 2-7$. {\it (iii) Setup III-Temperature and momentum dependent heavy quark transport coefficients:} For a heavy quark with a finite momentum, its dynamics in the QGP medium are described with parameters, $A(p, T)$, $B_0(p, T)$, $B_1(p, T)$ where,  in general, $B_0\ne B_1$ (see eqs.~\ref{eq:A} and~\ref{eq:B}). The momentum and temperature dependence of the heavy quark drag and diffusion coefficients is described in eqs.~\ref{1.610}-\ref{1.612}. It is seen that the drag coefficient decreases with an increase in heavy quark momentum, whereas the trend is quite the opposite for momentum diffusion coefficients. The transverse diffusion coefficient $B_0$ increases with momentum and saturates at higher momentum. However, the coefficient $B_1$ has a sharp rise with an increase in heavy quark momentum, which indicates large random kicks to the heavy quark. The fluctuation-dissipation relation is enforced for the longitudinal diffusion coefficient to ensure the equilibration of heavy quarks in the medium. The details of the derivation of drag and diffusion coefficient are given in appendix A. We have also analyzed the viscous effects to the heavy quark transport coefficients. It is seen that viscous effects have no visible impact on the temperature behavior of $D_s$, especially in the high-temperature regime~\cite{Kurian:2020orp}. 

In MARTINI, Peterson fragmentation is employed to describe the heavy quark hadronization process. For the open heavy flavor mesons, heavy quark fragments into a meson, and the fragmentation function estimates the fractional momentum of the resulting hadron. Quarkonium is another possible final state for the heavy quarks, and MARTINI separates out this final state from open heavy flavor mesons using a three-step algorithm as described in detail in~\cite{Young:2011ug}. 

%%%%%%%%%%%%%%%%%%%%%%%%%%%%%%%%%%%%%%%%%%%%%%%%%%%
%%%%%%%%%%%%%%%%%%%%%%%%%%%%%%%%%%%%%%%%%%%%%%%%%%%%

{\bf \emph{Results and discussions}}- 
\begin{figure}
    \includegraphics[width=0.45\textwidth]{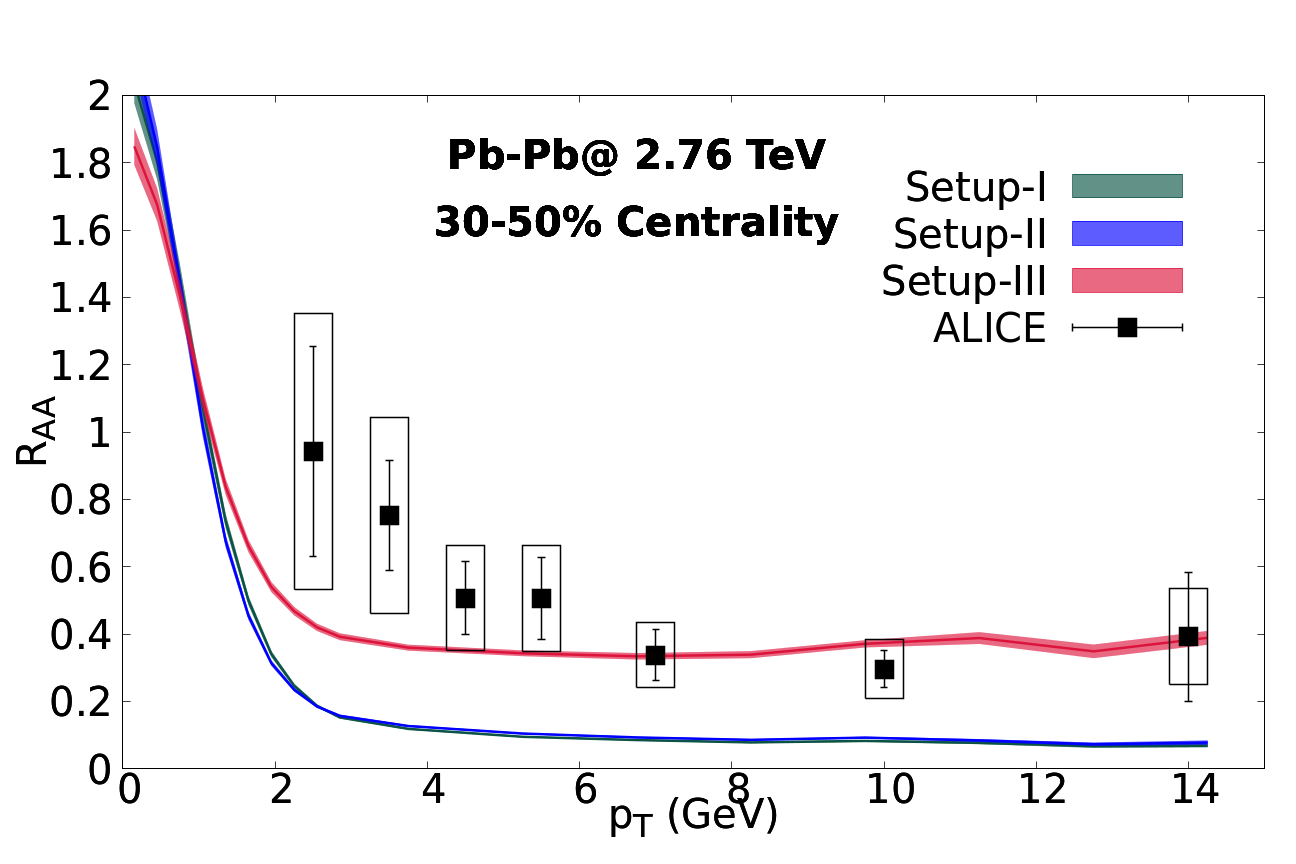}
    \includegraphics[width=0.45\textwidth]{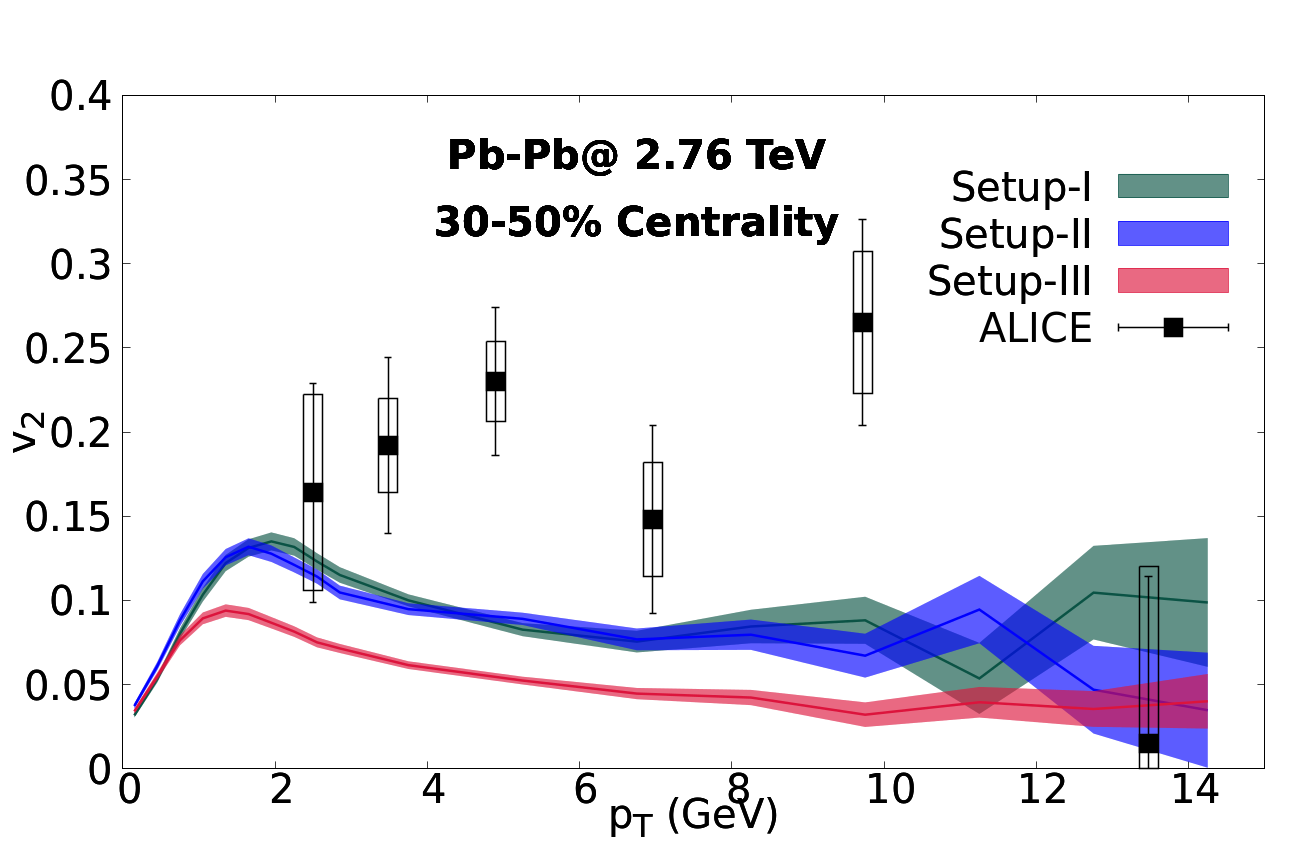}
    \caption{Nuclear suppression factor and elliptic flow of D mesons in three different setups. Experimental data of D meson $R_{AA}$ and $v_2$ are from the ALICE collaboration, Refs.~\cite{ALICE:2015vxz} and \cite{ALICE:2013olq}, respectively. The nuclear suppression factor for setups I and II are almost identical.}
    \label{fig2}
\end{figure}
We evaluate the $R_{AA}$ and $v_{2}$ of D mesons. We used 100 hydrodynamic events per 10\% centrality to simulate the collective motion. For every hydrodynamic event, 50000 different charm configurations were used. The elliptic flow harmonic $v_2$ was evaluated using the event plane method where the azimuthal angle of D meson $\phi_D$ is correlated with the second order event-plane angle $\Psi_2$. In experiments, the event plane angle is determined by the azimuthal distribution of all charged hadrons. We used the initial state spatial anisotropy to determine the event-plane angle. In the studied centrality class, the spatial anisotropy angle and the event plane angle are strongly correlated.

We compared the results with the available experimental data at $30-50\%$. The nuclear suppression factor $R_{AA}$  and elliptic flow $v_2$ of the D-mesons are shown in fig.~\ref{fig2}. We observe that the estimation with momentum and temperature-dependent charm quark transport coefficients (Setup-III) have a better agreement with the available data for $R_{AA}$ in comparison with the other two setups. In contrast, the setup underestimates the D meson $v_2$. A recent study~\cite{Das:2015ana} has predicted that the temperature dependence of heavy quark interaction strength plays a vital role in the simultaneous description of both $R_{AA}$ and $v_2$ of D meson at the RHIC energy. However, there are still mismatches between the calculations and measurements, especially at the LHC energies. Notably, none of the models could explain the enhancement in $v_2$ around $p_T = 10$ GeV. This could be due to the uncertainties of heavy quark interaction in the medium and heavy flavor hadronization process, especially in the low $p_T$ regime where the coalescence mechanism may have an impact on the observables~\cite{Fries:2008hs}.

 \begin{figure}
    \includegraphics[width=0.45\textwidth]{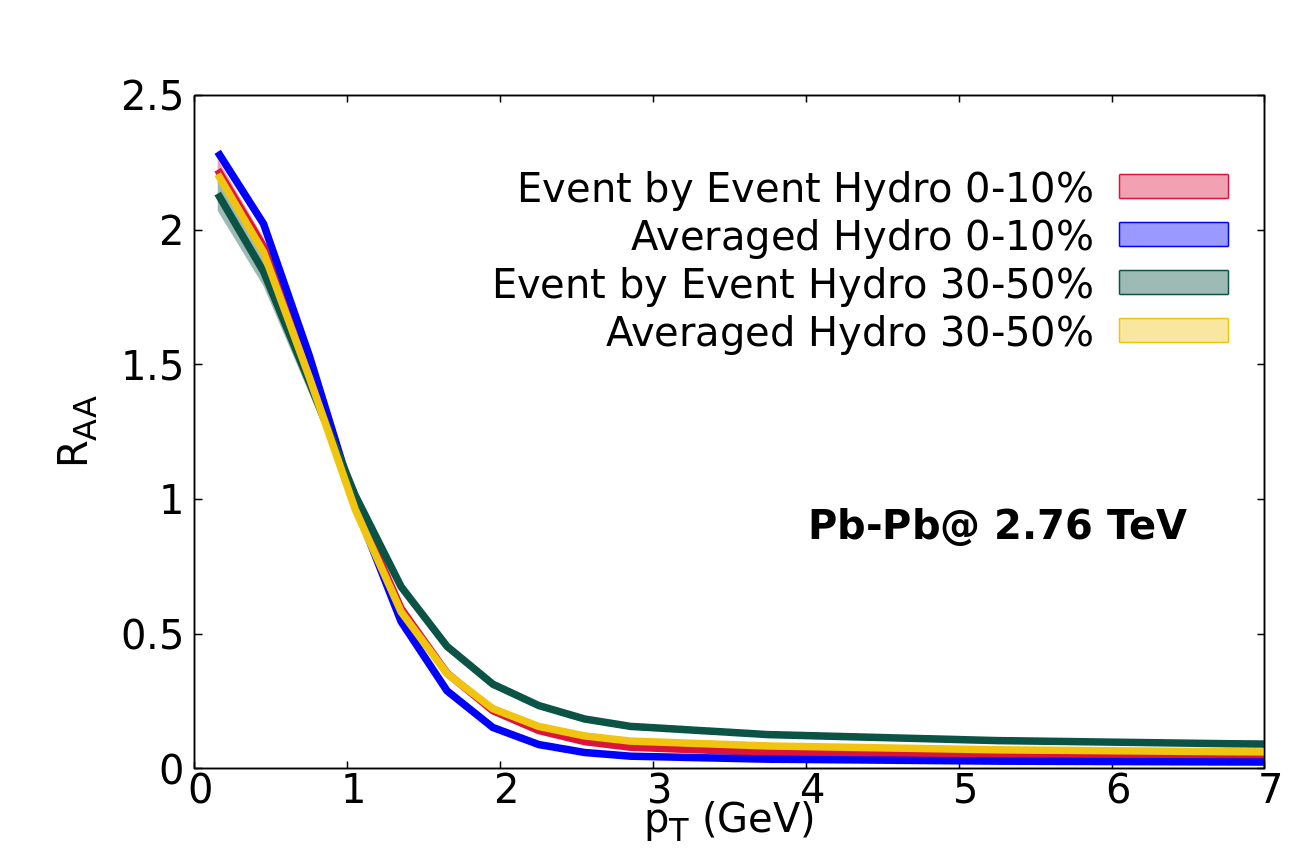}
    \includegraphics[width=0.45\textwidth]{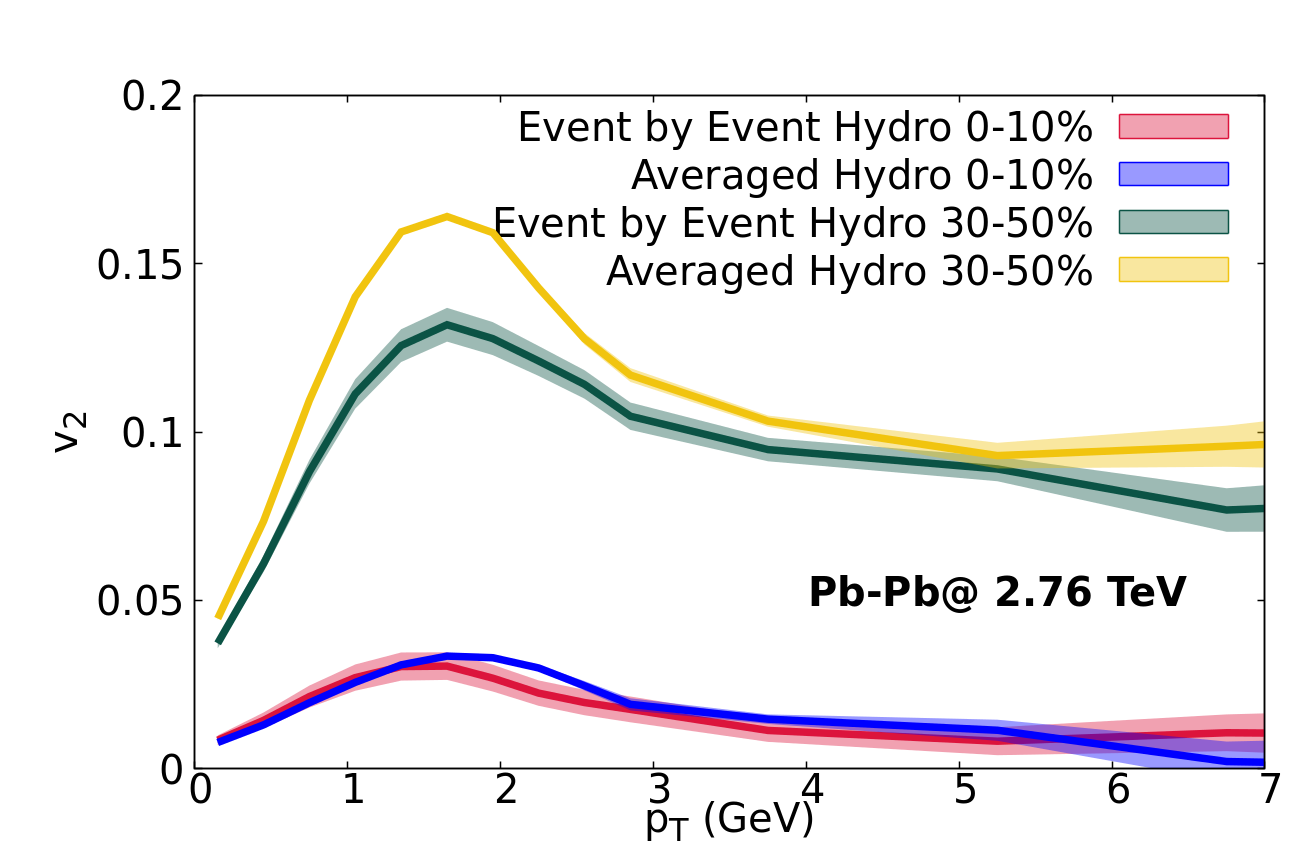}
    \caption{Nuclear suppression factor and elliptic flow of D mesons in setup-II in two different centralities with event-by-event initialized and smooth hydro backgrounds.}
    \label{fig3}
\end{figure}
 
To quantify the impact of fluctuating IP-Glasma initial state on heavy quark observables, we have compared the results from event-by-event initialized calculations with those from smooth initial conditions. The smooth initial profiles are obtained from the optical Glauber model for impact parameters 3.5 fm and 10 fm, which roughly correspond to the 0-10\% and 30-50\% centrality bins. All other parameters were held fixed.

The impact of fluctuating initial conditions on D meson $v_2$ is illustrated in fig.~\ref{fig3}. At low $p_T$, the fluctuating initial condition seems to have a significant influence on the D-meson $v_2$ for $30-50 \%$ centrality. The effect of fluctuating initial conditions can be understood as a convolution of two distinct effects. Fluctuations increase local pressure gradients and enhance flow. That will lead to an increase in $v_2$. However, fluctuations also increase decorrelation between the event planes of light flavor and heavy flavor mesons. As the two are produced by different mechanisms in different stages of evolution, their event plane angles are generally not identical. Heavy flavor meson $v_2$ is measured by taking its projection on the event plane determined by charged hadrons, which is dominated by the light flavor mesons. This increased decorrelation suppresses $v_2$. The net effect is a combination of two factors. As the charm quarks are much heavier than the background medium, the enhancement in $v_2$ from the increased flow is more than compensated by the decorrelation. This effect is opposite to that observed in jets~\cite{Noronha-Hostler:2016eow}.

{\bf \emph{Summary}}- In this paper, a hybrid framework is developed to study the evolution of heavy flavor by incorporating the recent developments in initial state dynamics and viscous QGP evolution in the relativistic heavy-ion collisions in Pb+Pb collision at $\sqrt{s_{NN}}= 2.76$ TeV. We introduce fluctuating IP-Glasma initial states and viscous hydrodynamics tuned to a global Bayesian analysis for the first time in a phenomenological study of the charm quark. The heavy quark dynamics is described within the Langevin approach in the expanding medium in which heavy quark coupling strength with the medium is quantified in terms of its transport coefficients. We explored the momentum and temperature dependence of the charm quark transport coefficients due to the collisional and radiative energy loss of the heavy quarks in the QGP medium, as well as its impact on the nuclear modification factor $R_{AA}$ and the elliptic flow $v_2$ of D-mesons. Our results with improved heavy quark dynamics with the latest developments in multistage hybrid frameworks for the dynamical evolution of collisions demonstrate that heavy flavor observables are influenced by the IP-Glasma initial state and bulk evolution of the medium. We see that fluctuating initial conditions have a significant effect on charm elliptic flow at low $p_T$. These effects are the result of an increase in both flow and decorrelations. While enhanced flow is the dominant effect for light jets, the event-plane decorrelation is more important for charm quarks. This indicates that the heavier charm quark is less susceptible to becoming part of the background flow than light quarks. Further, we observe that the energy loss profiles of a charm quark and non-perturbative effects in the QGP medium have a significant role in both $R_{AA}$ and $v_2$ of D-mesons. 

Looking into the future, it will be interesting to explore the influence of pre-equilibrium interactions on heavy quark energy loss. These effects are essential to maintain coherence in the theoretical description of heavy flavor dynamics in heavy-ion collisions~\cite{Das:2017dsh,Carrington:2022bnv,Boguslavski:2023fdm}. Additionally, it is an important aspect to take into account the uncertainties associated with the momentum and temperature dependence of heavy quark transport coefficients to simultaneously describe $R_{AA}$ and $v_{2}$ in Pb+Pb collisions at $2.76$ TeV. This tuning can be achieved by utilizing a model-to-data comparison. We leave these interesting aspects to the near future.

{\bf \emph{Acknowledgements}}- 
We thank Bj\"orn Schenke and Gojko Vujanovic for helpful discussions and feedback. We acknowledge Matthew Heffernan and Nicolas Fortier for their help with the IP-Glasma initial state files and with MUSIC parameters. Numerical computations were done on the resources provided by the Minnesota Supercomputing Institute (MSI) at the University of Minnesota and on Beluga supercomputer at McGill University managed by Calcul Qu\'ebec and the Digital Research Alliance of Canada. M.S. is supported by the U.S. DOE Grant No. DE-FG02-87ER40328. M.K. acknowledges a fellowship from the Fonds de recherche du
Qu\'ebec - Nature et technologies (FRQNT), support from the Natural Sciences and Engineering Research Council of Canada, and the Special Postdoctoral Researchers Program of RIKEN. S.J. and C.G. are supported by the Natural Sciences and Engineering Research Council of Canada under grant numbers SAPIN-2018-00024 and SAPIN-2020-00048 respectively.

%%%%%%%%%%%%%%%%%%%%%%%%%%%%%%%%%%%%%%%%%

\bibliography{ref}{}

%%%%%%%%%%%%%%%%%%%%%%%%%%%%%%%%%%%%%%%%%%%%%
 \appendix
 \section{Appendix A: temperature and momentum dependence of charm quark transport coefficients}
%%%%%%%%%%%%%%%%%%%%%%%%%%%%%%%%%%%%%%%%%%%
%%%%%%%%%%%%%%%%%%%%%%%%%%%%%%%%%%%%%%%
\begin{figure}[hbt!]
    \includegraphics[width=0.45\textwidth]{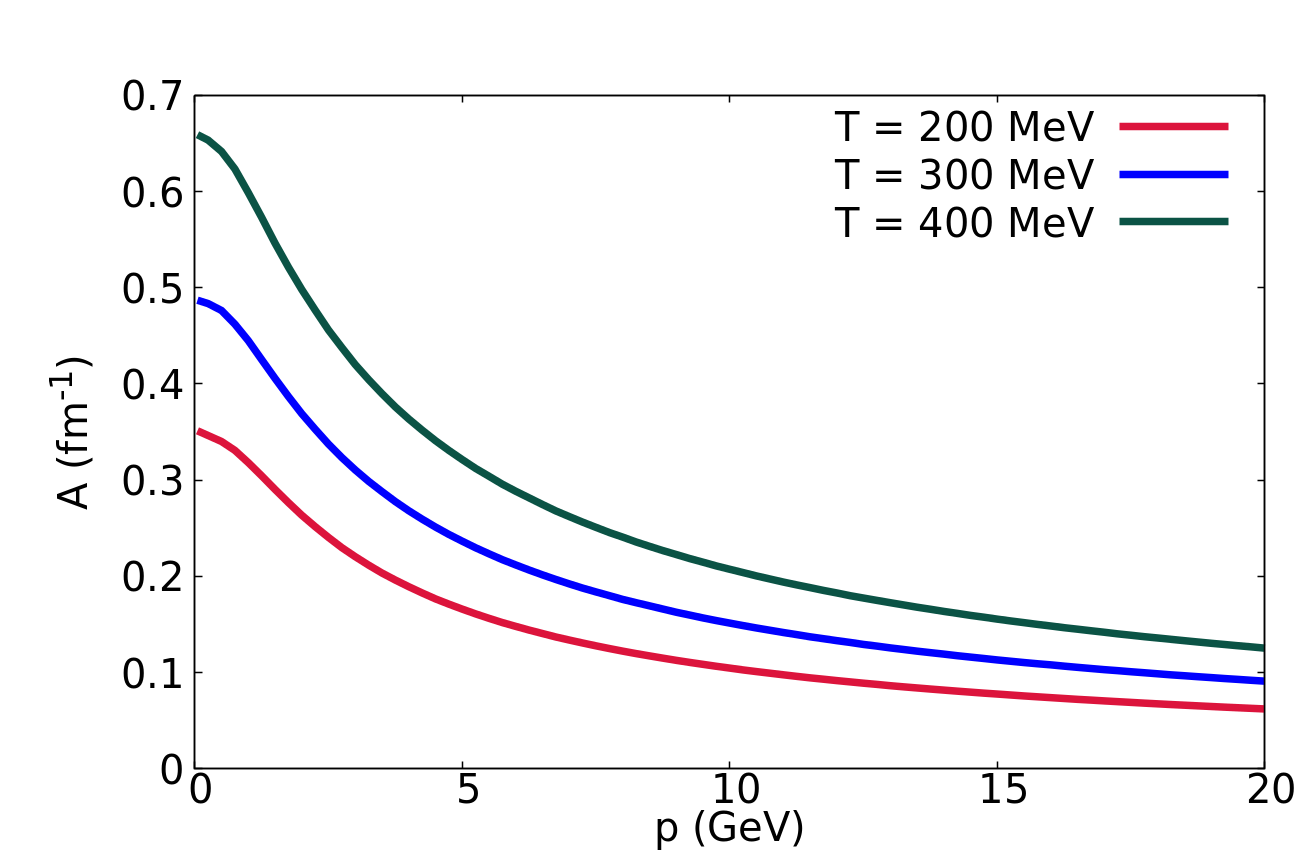}
    \includegraphics[width=0.45\textwidth]{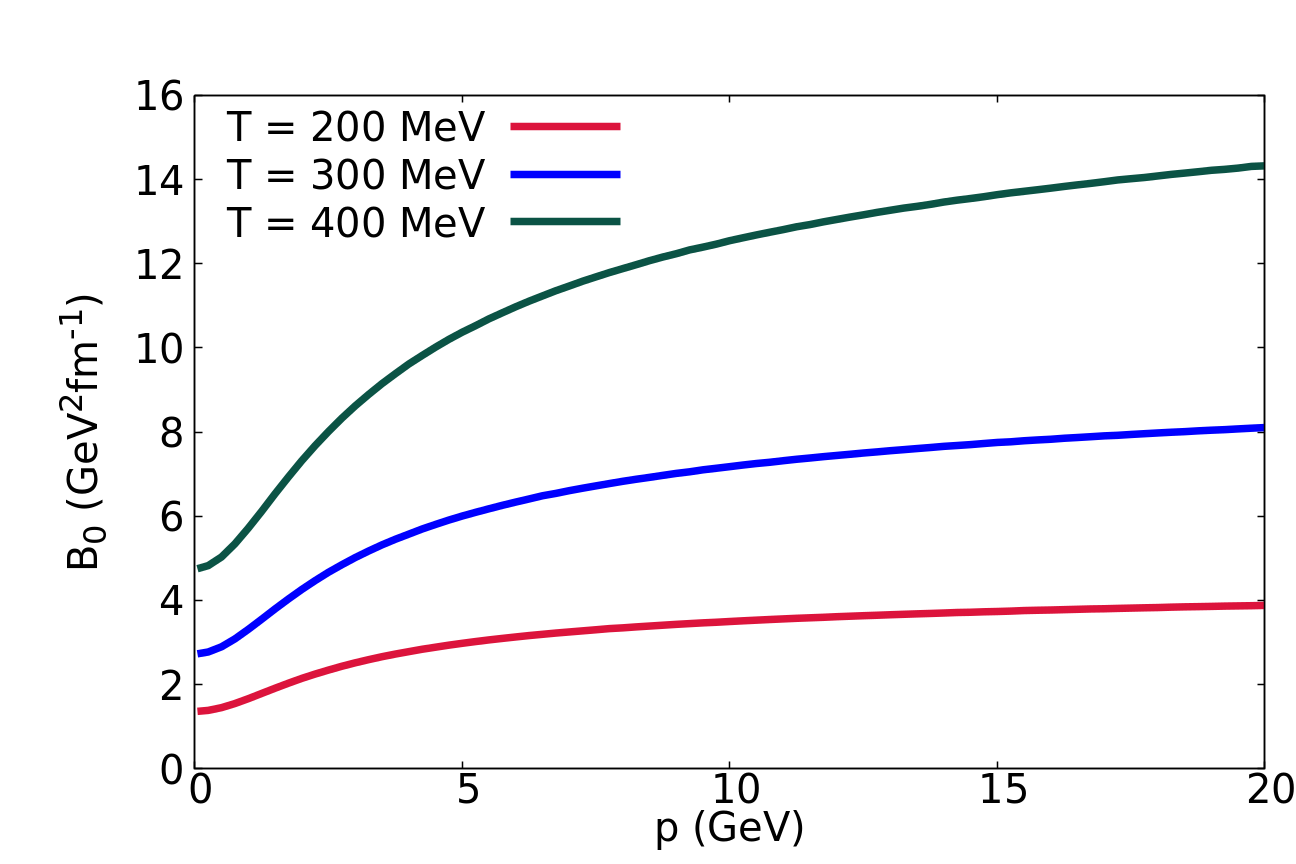}
    \caption{Drag coefficient A (top) and diffusion coefficient $B_0$ (bottom) as a function of charm quark momentum in the local rest frame.}
    \label{fig4}
\end{figure}
%%%%%%%%%%%%%%%%%%%%%%%%%%%%%%%%%%%%%%%%
By employing the Landau approximation~\cite{Landau}, the collision integral in the Boltzmann equation can be simplified and the transport coefficients can be defined as,
\begin{align}
&A=~\langle\langle1\rangle\rangle-\dfrac{\langle\langle{\bf p}\cdot {\bf p}'\rangle\rangle}{|{\bf p}|^2},\label{1.610}\\
&B_{0}=\dfrac{1}{4}\bigg[\langle\langle{|{\bf p}'|}^2\rangle\rangle-\dfrac{\langle\langle({\bf p}\cdot{\bf p}')^2\rangle\rangle}{|{\bf p}|^2}\bigg],\label{1.611}\\
&B_{1}=\dfrac{1}{2}\bigg[\dfrac{\langle\langle({\bf p}\cdot{\bf p}')^2\rangle\rangle}{|{\bf p}|^2}-2\langle\langle{\bf p}\cdot{\bf p}'\rangle\rangle+|{\bf p}|^2\langle\langle1\rangle\rangle\bigg]\label{1.612},
\end{align}
where $\langle\langle F(|{\bf p}'|)\rangle\rangle$ denote the thermal average of a function $F(|{\bf p}'|)$ and depends upon the heavy quark interaction process in medium.

For the elastic collisional process, $HQ(p)+l(q)\rightarrow HQ(p')+l(q')$, where $l$ denotes light quarks or gluons, $\langle\langle F(|{\bf p}'|)\rangle\rangle$ is defined as,
\begin{align}\label{Ap1c}
\langle\langle F(|{\bf p}'|)&\rangle\rangle=\dfrac{1}{\gamma_{HQ}}\dfrac{1}{2E_p}\int{\dfrac{d^3{\bf q}}{(2\pi)^3 2E_q}}\int \frac{d^3 {\bf p'}}{(2 \pi)^3 2E_{p'}}\nonumber\\
&\times\int \frac{d^3 {\bf q'}}{(2 \pi)^3 2E_{q'}}(2\pi)^4\delta^4(p+q-p'-q')\nonumber\\
&\times\sum|{\mathcal{M}}_{2\rightarrow 2}|^2 f_{g/q}({E_q})\Big(1 \pm f_{g/q}(E_{q'})\Big)F(|{\bf p}'|),
\end{align}
where $\gamma_{HQ}$ as the statistical degeneracy factor of heavy quark and $|{\mathcal{M}}_{2\rightarrow 2}|$ describes interaction amplitude of the heavy quark-thermal particles elastic scattering process. Here, $f_{g/q}$ is the distribution function of thermal particles in the evolving medium. 

For the inelastic ($2\rightarrow 3$) process,
$HQ(p)+l(q) \rightarrow HQ (p')+l(q') + g(k')$,
with $k'\equiv(E_{k'},{\bf k'_{\perp}},k'_z)$ as the four-momentum of the emitted soft gluon by the heavy quark in the final state, 
the thermal averaged $F(|{\bf p}'|)$ takes the form as~\cite{Mazumder:2013oaa},
\begin{align}\label{Ap1}
    \langle \langle F(|{\bf p}'|) &\rangle \rangle =\dfrac{1}{\gamma_{HQ}}\frac{1}{2 E_p} \int \frac{d^3 {\bf q}}{(2 \pi)^3 2E_q} \int \frac{d^3 {\bf p'}}{(2 \pi)^3 2E_{p'}} \nonumber\\ &\times \int \frac{d^3 {\bf q'}}{(2 \pi)^3 2E_{q'}} \int \frac{d^3 {\bf k'}}{(2 \pi)^3 2E_{k'}} \ (2 \pi)^4 \nonumber\\&\times \delta^{(4)}(p+q-p'-q'-k')  \sum{|{\mathcal{M}}_{2\rightarrow 3}|^2} \nonumber\\&\times f_{g/q}(E_q) (1 \pm f_{g/q}(E_{q'})) \ (1 + f_g(E_{k'})) \ \nonumber\\& \times \theta_1(E_p-E_{k'}) \ \theta_2(\tau-\tau_F) \ F(|{\bf p}'|),
\end{align}
where $|{\mathcal{M}}_{2\rightarrow 3}|^2$ describes the matrix element squared the radiative process~\cite{Abir:2011jb}. The theta function $\theta_1(E_p-E_{k'})$ imposes constraints on the heavy quark initial energy and $\theta_2(\tau-\tau_F)$ indicates that scattering time $\tau$ is larger than the gluon formation time $\tau_F$ (Landau-Pomeranchuk-Migdal Effect)~\cite{Gyulassy:1993hr,Klein:1998du}.

Fig.~\ref{fig4} shows the drag and diffusion coefficients of charm quark as a function of momentum while including the collisional and radiative processes with the modified IR regulator and effective coupling.

\end{document}